\documentclass[pdftex,twocolumn,epjc3]{svjour3}          

\RequirePackage[T1]{fontenc}

\usepackage[utf8]{inputenc}
\usepackage[english]{babel}
\usepackage{ragged2e,amsfonts,amssymb,xcolor}
\usepackage{amsmath}
\smartqed  

\RequirePackage{graphicx}
\RequirePackage{mathptmx}      
\RequirePackage{flushend}
\RequirePackage[numbers,sort&compress]{natbib}
\RequirePackage[colorlinks,citecolor=blue,urlcolor=blue,linkcolor=blue]{hyperref}

\journalname{Eur. Phys. J. C}

\begin{document}

\title{Consistency Conditions for Fields Localization on Braneworlds}


\author{L. F. F. Freitas\thanksref{e1,addr1}
        \and G. Alencar\thanksref{e2,addr1}
        \and R. R. Landim\thanksref{e3,addr1} 
}

\thankstext{e1}{e-mail: luizfreitas@fisica.ufc.br}
\thankstext{e2}{e-mail: geovamaciel@gmail.com}
\thankstext{e3}{e-mail: renan@fisica.ufc.br}

\institute{Universidade Federal do Ceará - UFC, Departamento de Física,\\ Campus do Pici - Bloco 922, Caixa Postal 6030, 60455-970 Fortaleza (CE) Brasil. \label{addr1}
}

\date{Received: date / Accepted: date}

\maketitle

\begin{abstract}
The general procedure for analyzing the localization of matter fields in Brane models is by integrating, in the action, its zero mode solutions over the extra dimensions. If this is finite, the field is said to be localized. However, the zero mode solutions must also satisfy the Einstein equations. With this in mind, we obtain stringent constraints on a general energy-momentum tensor by analyzing the Einstein's equations. These {\it consistency conditions} must be satisfied for any Braneworld model. We apply it for some fields of the Standard Model. For a free massless scalar field, the zero-mode localization is consistent only if the field does not depend on the extra dimensions. About the spin $\frac{1}{2}$ field with Yukawa-like interactions, we find a very specific relation between Yukawa function and the warp factor. As a consequence, the spinor field localization becomes inconsistent for most of the models studied in the literature. For the free vector field case, we find that the zero-mode do not satisfy the consistency conditions. Finally, we consider the mechanisms proposed to localize this field. We find that a few survive, and even for these, the consistency conditions fix the free parameters or the possible class of solutions allowed.
\end{abstract}

\section{Introduction}\label{ASec-1}

Braneworld scenarios gained prominence after the emergence of the $5$D warped models introduced by Randall-Sundrum (RS) \cite{RS1, RS2}. Since then, many other models with localized gravity have been proposed. Some of them also in $5$D, such as: thick brane generated by scalar fields with different potential functions \cite{Gremm01,Bazeia04}; deformed brane models with internal structure \cite{Bazeia05,Bazeia02}; thick brane with purely geometric features \cite{Cendejas}; beyond proposals in cosmological contexts \cite{Singh}; or $f(R)$ theories \cite{Hoff}, among others. A comprehensive and more detailed review of thick braneworld can be found in Ref. \cite{Dzhunushaliev}. Besides these $5$D scenarios, other proposals in higher-dimensional configuration were presented. For example, in $6-$dimension, braneworld models generated by string-like topological defect with a scalar field \cite{Gregory,Gherghetta}; vortex defect in the context of abelian Higgs model \cite{Giovannini01}; or cigar-like thick braneworld \cite{Carlos,Silva}. And also, versions of these models in cosmological context \cite{Cline,Kofinas}, among others \cite{Kim,Coley,Choudhury01}. There are also proposals in higher-dimensional scenarios \cite{Arkani02,Oda01,Bazeia01,Navarro01}. Amid this large variety of works, we can find some with the focus on the general features of braneworlds, as the Refs. \cite{Csaba,Gibbons}. In these papers are studied the consistency of the gravitational field, Newton's law on the brane, search for resonant gravitational modes, and other related issues.

In this braneworld context, beyond the gravitational field, the issue of the Standard Model (SM) fields localization should also be verified. Some general studies have been performed in the literature. The Ref. \cite{Barut} presents a detailed study of the free spin $0,\ \frac{1}{2}$ and $1$ fields localization in RS-II delta-like brane model. Among these fields only the free scalar field and the left-handed spinor can be confined on that model. In Refs. \cite{Oda02,Liu01,Liu03,Liu04} was analyzed the localization of the above fields for thick branes embedded in $AdS_{5}$ space. For these cases, the scalar and the left-handed spinor can be confined, and again the $U(1)$ gauge field is not localized. Another study, performed for thick branes embedded in $dS_{5}$ space \cite{Liu02,Guo}, showed the same results for the scalar and the spinor fields, however, unlike the early models, the free vector field can be confined in such models. This same analysis was also performed for other dimensional configurations. In $6$D string-like models, for example, the Refs. \cite{Oda03,Oda06} show that the free spin $0$ and $1$ fields can be localized. In Ref. \cite{Midodashvili}, the author shows that, beyond the scalar and the $U(1)$ vector fields, the free spin $\frac{1}{2}$ field can also be confined on the brane. In arbitrary higher-dimensions, the Refs. \cite{Oda04,Oda05} obtained the same results above for the Standard Model fields. With this, we already have an indication that the confinement of fields, mainly the spinor and the gauge vector fields, is closely related to the geometric features of space. In another direction, aiming to obtain the localization of fields, some mechanisms were proposed. For example, in Refs. \cite{Koley01,Ringeval,Melfo,Castro,Mendes} the localization of spin $\frac{1}{2}$ field is obtained for various braneworld scenarios by proposing a Yukawa-like coupling. The Ref. \cite{Batell} provides the confinement of a non-abelian Yang-Mills field by introducing non-minimal couplings with gravity. Among others, for these and other fields, for those and various other braneworld models \cite{Casadio,Alencar01,Germani,GLandim01,GLandim02}.

In most $5-$dimensional braneworlds, the free $U(1)$ gauge field cannot be confined, as mentioned above. However, by adding some suitable interaction terms, that field can be localized. For example, in RS-II like models, the localization can be obtained by adding an interaction term between a $3$-form field (topological term) and the vector field \cite{Oda07}; or by proposing a non-covariant mechanism with two mass terms, one in the bulk and another just in brane \cite{Ghoroku}, the confinement is obtained by fine-tuning these mass parameters; or still, by proposing non-minimal interactions between the gravity and the vector field through the Ricci scalar \cite{Alencar02,Zhao}. For $5$D thick brane models, the Refs. \cite{Kehagias,Fu01,Chumbes,Alencar03,Cruz01,Cruz02} proposed a modified kinetic term of the vector gauge field by adding couplings with a scalar field function. In $6$D models, the Ref. \cite{Flachi} proposed a delta-like brane generated by brane intersection and, to confine the vector field, the author proposes interaction terms between this and the scalar Ricci, and/or the Ricci tensor. In higher-dimensional models, the Ref. \cite{Freitas} carried out a general study of the abelian vector field localization through the couplings with the scalar and the Ricci tensor.

Despite these various results indicate that the Standard Model fields can be well-defined on the braneworld scenarios, there is not yet a study on the consistency of the localization procedure. Generally, when we talk about fields localization on braneworlds, it means that we wish to factor out an action $S^{(D)}$ defined on the bulk into a sector containing an effective action $S^{(4)}_{eff}$ on the $3$-brane and an integral $K$ in the coordinates of the extra dimensions. Thus, we say that the theory is well-defined, {\it i.e.}, the field is localized on the brane, when the integral $K$ is finite. In this manuscript, we aim to analyze the consistency of this localization procedure with the Einstein's equations for the SM fields. Special attention is paid to the $U(1)$ gauge field, both the free field and the cases where some localization mechanism is used. This work is organized as follows: In section (\ref{ASec-2}), we obtain two general consistency conditions that must be satisfied by any field in order that the localization procedure to be consistent. We apply these conditions for spin $0$ and $\frac{1}{2}$ fields in sections (\ref{ASec-3}) and (\ref{ASec-4}), respectively. In section (\ref{ASec-5}), we carry out the consistency analysis for the vector field, free and with some localization mechanism. Conclusions are left for section (\ref{ASec-6}).

\section{Einstein Equations - Consistency Conditions}\label{ASec-2}

To perform a more general and comprehensive discussion, let us consider a braneworld in $(D=d+n)-$dimensions with metric given by the generic ansatz
\begin{eqnarray}\label{2-01}
ds^{2}\!=\!g_{MN}dx^{N}\!dx^{M}\!=e^{2\sigma(y)}\hat{g}_{\mu\nu}(x)dx^{\mu}\!dx^{\nu}\!+\Bar{g}_{jk}(y)dy^{j}dy^{k}.
\end{eqnarray}
Where $\underline{d}$ is the brane dimension, indexed by $(\mu,\nu,..)$, and $\underline{n}$ is the number of extra dimensions, labeled by $(j,k,..)$. Beyond this, the metric (\ref{2-01}) will be considered diagonal with signature $(-,+,+,..)$.

As mentioned before, in the study of fields localization on braneworld scenarios like (\ref{2-01}), we wish to factor out a matter Lagrangian like
\begin{eqnarray}\label{2-02}
S=\int d^{d}xd^{n}y\sqrt{-g^{(D)}}\mathcal{L}^{(D)}(x,y),
\end{eqnarray}
into an effective action on the brane and an integral in the extra dimension, i.e., 
\begin{eqnarray}\label{2-03}
S&=&\int d^{n}yf(y)\int d^{d}x\sqrt{-\hat{g}^{d}(x)}\mathcal{L}_{ef}^{(d)}(x)\nonumber\\&=&K\int d^{d}x\sqrt{-\hat{g}^{d}(x)}\mathcal{L}_{ef}^{(d)}(x).
\end{eqnarray}
From this, we say that the theory is well-defined, or the field is localized, if $K$ is finite. In doing this, the metric (\ref{2-01}) is considered only as a background previously determined by some process. And also, it is assumed that the matter Lagrangian (\ref{2-02}) will not change the background geometry. Here, we will discuss exactly the consistency of this last assumption by studying the Einstein's equations.

The Einstein-Hilbert action for an arbitrary braneworld model, with a matter Lagrangian like (\ref{2-02}), can be written as 
\begin{eqnarray}\label{2-04}
S&=&S_{brane}+S_{matter}\nonumber\\
&=&\!\int\! d^{d}xd^{n}y\sqrt{-g}\!\left[\!\frac{1}{2\kappa^{2}}\left(R-2\Lambda\right)\!+\!\mathcal{L}^{b}(y)\!+\!\mathcal{L}^{m}(x,y)\right].
\end{eqnarray}
Here, $\mathcal{L}^{b}$ is related to the brane generation mechanism, and it is a function only of extra dimensions. The term $\mathcal{L}^{m}$ is the matter Lagrangian related to the Standard Model fields, which the confinement should be studied. By performing the variation of action (\ref{2-04}) with respect to the metric, we get the bellow Einstein's equations
\begin{eqnarray}\label{2-05}
R_{MN}-\frac{1}{2}g_{MN}R+g_{MN}\Lambda=\kappa^{2}\left(T^{b}_{MN}+T^{m}_{MN}\right).
\end{eqnarray}
We should add to Eq. (\ref{2-05}) the equations of motion (EOM) related to the fields in the Lagrangians $\mathcal{L}^{b}$, and $\mathcal{L}^{m}$. Fortunately, these EOM's are not important to our discussion, and they will not be written here. From metric (\ref{2-01}), we can get the Ricci tensor components,
\begin{eqnarray}
R_{\mu\nu}(x,y)&=&\hat{R}_{\mu\nu}(x)-\frac{1}{d}\hat{g}_{\mu\nu}(x)e^{-(d-2)\sigma(y)}\nabla_{k}\nabla^{k}e^{d\sigma(y)},\label{2-06}\\
R_{jk}(x,y)&=&\Bar{R}_{jk}(y)-de^{-\sigma(y)}\nabla_{j}\nabla_{k}e^{\sigma(y)},\label{2-07}
\end{eqnarray}
and the components $R_{j\mu}=0$. From these, the Ricci scalar can be written as
\begin{eqnarray}
R(x,y)=\hat{R}(x)e^{-2\sigma}+\Bar{R}(y)&-&de^{-\sigma(y)}\nabla_{k}\nabla^{k}e^{\sigma(y)}\nonumber\\&-&e^{-d\sigma(y)}\nabla_{k}\nabla^{k}e^{d\sigma(y)},\label{2-08}
\end{eqnarray}
where $\Bar{R}=\Bar{g}^{jk}\Bar{R}_{jk}$ and $\hat{R}=\hat{g}^{\mu\nu}\hat{R}_{\mu\nu}$. That way, the Eq. (\ref{2-05}) can be separated in the following two equations 
\begin{eqnarray}
G_{\mu\nu}(x,y)+g_{\mu\nu}(x,y)\Lambda=\kappa^{2}\left[T^{b}_{\mu\nu}(x,y)+T^{m}_{\mu\nu}(x,y)\right],\label{2-09}\\
G_{jk}(x,y)+g_{jk}(x,y)\Lambda=\kappa^{2}\left[T^{b}_{jk}(x,y)+T^{m}_{jk}(x,y)\right].\label{2-10}
\end{eqnarray}
By using Eqs. (\ref{2-06}), (\ref{2-07}) and (\ref{2-08}), the equation (\ref{2-09}) can be written as
\begin{eqnarray}\label{2-11}
\hat{G}_{\mu\nu}(x)+\hat{g}_{\mu\nu}(x)\!\left[\!\frac{1}{2}S(y)-\frac{1}{2}\Bar{R}(y)+\Lambda\!\right]\!e^{2\sigma(y)}\nonumber\\=\kappa^{2}\!\left[T^{b}_{\mu\nu}(x,y)+T^{m}_{\mu\nu}(x,y)\right],
\end{eqnarray}
where $S(y)=(d-1)\left[2\nabla_{k}\nabla^{k}\sigma+d\nabla^{k}\sigma\nabla_{k}\sigma\right]$. When the equation (\ref{2-11}) is solved to obtain the vacuum braneworld metric, the matter Lagrangian $\mathcal{L}^{m}$ is considered equal to zero, and the Lagrangian $\mathcal{L}^{b}$ is a function only of the extra dimensions. In this setup, $T^{b}_{\mu\nu}(x,y)=\hat{g}_{\mu\nu}(x)e^{2\sigma(y)}\mathcal{L}^{b}(y)$, and, with this, we can perform the separation of variables in Eq. (\ref{2-11}) as 
\begin{eqnarray}\label{2-12}
\left[\frac{1}{2}S(y)-\frac{1}{2}\Bar{R}(y)+\Lambda-\kappa^{2}\mathcal{L}^{b}(y)\right]\!e^{2\sigma(y)}=\alpha,
\end{eqnarray}
with $\alpha$ a constant that will be interpreted as the cosmological constant on the brane. Thus, braneworld metric in the vacuum $(\mathcal{L}^{m}=0)$ should satisfy the equation (\ref{2-12}). In the study of fields localization, where the starting point is the matter Lagrangian $\mathcal{L}^{m}$, the metric is exactly that obtained in the vacuum, and therefore it satisfies (\ref{2-12}). Now, let us assume that Eq. (\ref{2-12}) is still valid even after adding Lagrangian $\mathcal{L}^{m}$. With this, Eq. (\ref{2-11}) can be written as,
\begin{eqnarray}
\hat{G}_{\mu\nu}(x)+\hat{g}_{\mu\nu}(x)\alpha=\kappa^{2}T^{m}_{\mu\nu}(x,y). \label{2-13}
\end{eqnarray}
Thus, we observe that the left-hand side of (\ref{2-13}) does not depend on the extra dimensions $y^{j}$, therefore, for consistency reasons, the energy-momentum tensor of matter field should satisfy the following condition 
\begin{eqnarray}\label{2-14}
T^{m}_{\mu\nu}(x,y)=\hat{T}^{m}_{\mu\nu}(x).
\end{eqnarray}
About equation (\ref{2-10}), it can also be simplified by using Eqs. (\ref{2-06}), (\ref{2-07}) and (\ref{2-08}). By doing this, we get a condition on the components $T^{m}_{jk}$ given by
\begin{eqnarray}\label{2-15}
 T^{m}_{jk}(x,y)=\left\lbrace\begin{array}{c}
 0, \ \mbox{or} \\ 
 \Bar{g}_{jk}(y)e^{-2\sigma(y)}\Bar{T}^{m}(x).
 \end{array} \right.
\end{eqnarray}
These consistency conditions, (\ref{2-14}) and (\ref{2-15}), are completely independent of the brane model, the number of extra dimensions, and also of the matter field considered in $\mathcal{L}^{m}$. That way, such conditions have a general valid, and should be satisfied for any model with the features considered above. Note that these conclusions are closely related to the possibility of the metric not changing by the presence of matter fields. In other words, these consistency conditions mean that back-reaction effects from the matter fields on the bulk geometry can be eliminated. Let us apply these results for some known cases.

\section{Applications - Scalar field}\label{ASec-3}

Let us start by discussing the scalar field localization. The Lagrangian for a massless scalar field in the braneworld model (\ref{2-01}) is given by
\begin{eqnarray}\label{3-01}
\mathcal{L}^{m}(x,y)=-\frac{1}{2}\partial_{M}\Phi(x,y)\partial^{M}\Phi(x,y).
\end{eqnarray} 
In studying the localization of this field, we can obtain the equation of motion
\begin{eqnarray}\label{3-02}
e^{(d-2)\sigma}\sqrt{\Bar{g}}\partial_{\mu}\left[\sqrt{-\hat{g}}\hat{g}^{\mu\nu}\partial_{\nu}\Phi\right]\ \ \ \ \ \ \ \ \ \ \ \ \ \ \nonumber\\+\sqrt{-\hat{g}}\partial_{j}\left[\sqrt{\Bar{g}}e^{d\sigma}\Bar{g}^{jk}\partial_{k}\Phi\right]=0,
\end{eqnarray}
where $\Bar{g}$ is the determinant of $\Bar{g}_{jk}(y)$, and $\hat{g}$ is the determinant of $\hat{g}_{\mu\nu}(x)$. With this, by proposing $\Phi(x,y)=\phi(x)\xi(y)$, it is possible to separate the variables for the zero-mode as
\begin{eqnarray}
\frac{1}{\sqrt{-\hat{g}}}\partial_{\mu}\left[\sqrt{-\hat{g}}\hat{g}^{\mu\nu}\partial_{\nu}\phi_{0}(x)\right]=0,\label{3-03}\\
-\frac{e^{-(d-2)\sigma}}{\sqrt{\Bar{g}}}\partial_{j}\left[\sqrt{\Bar{g}}e^{d\sigma}\Bar{g}^{jk}\partial_{k}\xi_{0}(y)\right]=0.\label{3-04}
\end{eqnarray}
From this, a solution for (\ref{3-04}) can be obtained, and the localization can be analyzed. As discussed in many Refs. \cite{Barut,Kehagias,Oda02,Oda03,Liu01,Liu02,Fu01,Midodashvili}, there is a constant solution for (\ref{3-04}) that can be confined for a wide variety of models. In order to test the consistency conditions (\ref{2-14}) and (\ref{2-15}) for zero-mode, let us calculate the energy-momentum tensor from (\ref{3-01}). In doing this, we get
\begin{eqnarray}\label{3-05}
T^{m}_{MN}(x,y)&=&-\frac{2}{\sqrt{-g}}\frac{\delta\left(\sqrt{-g}\mathcal{L}^{m}\right)}{\delta g^{MN}}\nonumber\\
&=&\partial_{M}\Phi(x,y)\partial_{N}\Phi(x,y)+g_{MN}\mathcal{L}^{m}(x,y).
\end{eqnarray}
By using the constant solution $\xi_{0}(y)=c_{0}$ for (\ref{3-04}), the components $T^{m}_{\mu\nu}$, obtained from equation (\ref{3-05}), can be written as 
\begin{eqnarray}\label{3-06}
T^{m}_{\mu\nu}(x,y)=c^{2}_{0}\left[\partial_{\mu}\phi(x)\partial_{\nu}\phi(x)+\hat{g}_{\mu\nu}(x)\hat{L}_{0}^{m}(x)\right].
\end{eqnarray}
Therefore, the consistency condition (\ref{2-14}) is immediately satisfied. About condition (\ref{2-15}), we can get, from (\ref{3-05}), for the zero-mode, that  
\begin{eqnarray}\label{3-07}
T^{m}_{jk}(x,y)=\Bar{g}_{jk}(y)e^{-2\sigma(y)}c^{2}_{0}\hat{L}_{0}^{m}(x).
\end{eqnarray}
Thus, by comparing this with (\ref{2-15}), we conclude that both consistency conditions are satisfied. Therefore, the free scalar field (zero-mode) localization is consistent with Einstein's equations, and any possible back-reaction effect from the scalar field on the background metric must be caused by the massive modes. Note that nowhere was it necessary to define the braneworld model, or the number of extra dimensions, for the consistency conditions to be met. In this way, these results for the zero-mode of scalar field are valid for a wide variety of models, whether for those with thin or thick brane, and for arbitrary codimension. 

\section{Applications - Spinor field}\label{ASec-4}

Now let us see briefly the spin $\frac{1}{2}$ field localization for an arbitrary codimension $1$ model. In this particular configuration, the metric (\ref{2-01}) will be written as
\begin{equation}\label{3-08}
ds^{2}=e^{2\sigma_{1}(y)}\hat{g}_{\mu\nu}(x)dx^{\mu}dx^{\nu}+e^{2\sigma_{2}(y)}dy^{2},
\end{equation}
with $\mu,\nu=1,2,...,d$ (even). Beyond this, the consistency conditions are given by 
\begin{eqnarray}\label{3-09}
T^{m}_{\mu\nu}(x,y)=\hat{T}^{m}_{\mu\nu}(x), \ \ \ \mbox{and}\ \ \ T^{m}_{jk}(x,y)=\left\lbrace\begin{array}{c}
 0, \ \mbox{or} \\ 
 e^{2(\sigma_{2}-\sigma_{1})}\Bar{T}^{m}(x).
 \end{array} \right.
\end{eqnarray}
We will consider the spinor field coupled to an arbitrary scalar function $f(y)$ through a Yukawa-like interaction term. In order that the Lagrangian for this case will be written as
\begin{eqnarray}\label{3-10}
\mathcal{L}^{m}(x,y)=-i\Bar{\Psi}\Gamma^{M}D_{M}\Psi+\lambda f(y)\Bar{\Psi}\Psi,
\end{eqnarray}
where $D_{M}=\partial_{M}+\omega_{M}$, and $\omega_{M}=\frac{1}{4}\omega^{ab}_{M}\Gamma_{a}\Gamma_{b}$ are the spin connections. The Gamma matrix in curved space $\Gamma^{M}$ are related to those in a local flat frame by\footnote{Here, the index $M,N,..$ are related to the curved space, and $a,b,..$ are related to the flat space. Beyond this, the vierbein satisfies $E^{M}_{a}E^{Na}=g^{MN}$, and the Gamma matrix satisfy $\left\lbrace\Gamma^{M},\Gamma^{N}\right\rbrace=-2g^{MN}$.} $\Gamma^{M}(x,y)=E^{M}_{a}(x,y)\Gamma^{a}$. From (\ref{3-10}), the following equation of motion can be obtained,
\begin{eqnarray}\label{3-11}
\left[i\Gamma^{M}D_{M}-\lambda f(y)\right]\Psi=0.
\end{eqnarray}
By defining\footnote{The vierbein $\hat{e}^{\mu}_{a}(x)$ should satisfy $\hat{e}^{\mu}_{a}(x)\hat{e}^{\nu a}(x)=\hat{g}^{\mu\nu}(x)$.} $E^{\mu}_{a}(x,y)=e^{-\sigma_{1}}\hat{e}^{\mu}_{a}(x)$, $E^{y}_{a}=0$ ($a=1,2,..,d$), $E^{\mu}_{d+1}=0$ and $E^{y}_{d+1}=e^{-\sigma_{2}}\delta^{y}_{d+1}$, the spin connections can be calculated
\begin{eqnarray}\label{3-12}
\omega_{\mu}(x,y)=\hat{\omega}_{\mu}(x)+\frac{1}{2}\partial_{y}\sigma_{1}\Gamma_{\mu}\Gamma^{y},\ \ \ \omega_{y}(x,y)=0.
\end{eqnarray}
And with this, the equation (\ref{3-11}) gives us
\begin{eqnarray}\label{3-13}
i\hat{\Gamma}^{\mu}(x)\hat{D}_{\mu}\Psi-i\frac{d}{2}e^{\sigma_{1}}\partial_{y}\sigma_{1}\Gamma^{y}\Psi &+&ie^{\sigma_{1}}\Gamma^{y}\partial_{y}\Psi\nonumber\\&-&\lambda e^{\sigma_{1}}f(y)\Psi=0,
\end{eqnarray}
where $\hat{D}_{\mu}=\partial_{\mu}+\hat{\omega}_{\mu}(x)$. Here, to solve the above equation, let us consider that the zero-mode solution satisfies $-i\hat{\Gamma}^{\mu}\hat{D}_{\mu}\Psi_{0}=0$. That way, we get for the massless mode
\begin{eqnarray}\label{3-14}
i\frac{d}{2}e^{\sigma_{1}-\sigma_{2}}\partial_{y}\sigma_{1}\Gamma^{d+1}\Psi_{0}&-&ie^{\sigma_{1}-\sigma_{2}}\Gamma^{d+1}\partial_{y}\Psi_{0}\nonumber\\&+&\lambda e^{\sigma_{1}}f(y)\Psi_{0}=0.
\end{eqnarray}
At this point, we will use a Gamma matrix representation such that, $\Gamma^{d+1}$ is a $d\times d$ diagonal matrix (remember that $d$ is even) in the following shape
\begin{eqnarray}\label{3-15}
\Gamma^{d+1}=i\left[\begin{array}{cc}
\mathbf{I}_{\frac{d}{2}} & \ \mathbf{0}_{\frac{d}{2}} \\ 
\mathbf{0}_{\frac{d}{2}} & -\mathbf{I}_{\frac{d}{2}}
\end{array}\right],
\end{eqnarray}
where $\mathbf{I}_{\frac{d}{2}}$ is a $\frac{d}{2}\times\frac{d}{2}$ identity matrix. Thus, we can define the $\frac{d}{2}$-dimensional spinors $\Psi_{0}^{\pm}(x,y)$, such that,
\begin{eqnarray}\label{3-16}
\Psi_{0}(x,y)= \left[\begin{array}{c}
\Psi_{0}^{+}(x,y) \\ 
\Psi_{0}^{-}(x,y)
\end{array}\right]= \left[\begin{array}{c}
\psi_{0}^{+}(x)\xi^{+}(y) \\ 
\psi_{0}^{-}(x)\xi^{-}(y)
\end{array}\right].
\end{eqnarray}  
And, Eq. (\ref{3-14}) can be split as 
\begin{eqnarray}\label{3-17}
\pm \frac{d}{2}\partial_{y}\sigma_{1}\xi^{\pm}(y)\mp \partial_{y}\xi^{\pm}(y)-\lambda e^{\sigma_{2}}f(y)\xi^{\pm}(y)=0.
\end{eqnarray}
The zero-mode solutions for (\ref{3-17}) are given by
\begin{eqnarray}
\xi^{+}(y)&=&c_{1}e^{-\lambda \int_{y} dy'e^{\sigma_{2}(y')}f(y')+\frac{d}{2}\sigma_{1}},\label{3-18}\\
\xi^{-}(y)&=&c_{2}e^{\lambda \int_{y} dy'e^{\sigma_{2}(y')}f(y')+\frac{d}{2}\sigma_{1}}.\label{3-19}
\end{eqnarray}
Therefore, by specifying the function $f(y)$, and the warp factors $\sigma_{1}$ and $\sigma_{2}$, the localization discussion can be performed. As we are interested in verify the consistency conditions (\ref{3-09}), let us calculate the energy-momentum tensor for these zero-mode solutions. From Lagrangian (\ref{3-10}), we get the components
\begin{eqnarray}\label{3-20}
\mathcal{T}^{m}_{\mu\nu}(x,y)=e^{\sigma_{1}(y)}\xi^{2}_{0}\left[\Bar{\psi}\hat{\Gamma}_{(\mu}\hat{D}_{\nu)}\psi+\hat{g}_{\mu\nu}(x)\hat{L}_{0}^{m}(x)\right].
\end{eqnarray}
Then, the first consistency condition in (\ref{3-09}) is satisfied if $e^{\sigma_{1}(y)}\xi^{2}_{0}(y)$ is a constant quantity. By using the solutions (\ref{3-18}) and (\ref{3-19}), we get 
\begin{eqnarray}\label{3-21}
\pm 2\lambda \int_{y} dy'e^{\sigma_{2}(y')}f(y')+(d+1)\sigma_{1}(y)&=&\kappa_{\pm},\ \mbox{or}\nonumber\\
\pm 2\lambda e^{\sigma_{2}(y)}f(y)+(d+1)\sigma_{1}'(y)&=&0.
\end{eqnarray}
Here, $\kappa_{\pm}$ are constants, and prime is the derivative with respect to the extra dimension. These relations in (\ref{3-21}) should be valid for any value of $y$. Note that, already for the free case ($\lambda=0$), consistency cannot be obtained. In fact, we should have $\sigma'(y)=0$, and this is not satisfied for any non-factorizable braneworld model. For models with $\lambda\neq 0$, we conclude that the spinor field localization can be made consistent with Einstein's equations only if $f(y)\propto e^{-\sigma_{2}(y)}\sigma_{1}'(y)$. When we analyze some models in the literature, the mechanism presented in Refs. \cite{Barut, Oda04}, for RS-II model \cite{RS2}, can be made consistent. However, for thick brane models, the localization mechanisms presented in Refs. \cite{Koley01,Kehagias,Melfo,Castro,Mendes,Ringeval} are not consistent, in such way those mechanisms should be reviewed. Otherwise, the braneworld metric should be modified to take into account the presence of the spinor field.

\section{Applications - Vector field}\label{ASec-5}

Now, we will discuss the consistency of the vector field localization. This subject was already treated early in the literature for codimension 1 delta-like models \cite{Duff}. In such reference, the authors show that the free vector field localization (zero-mode) is not consistent with Einstein's equations. Here, let us discuss this issue for the free field in an arbitrary braneworld and also for some localization mechanisms commonly used in the literature.

\subsection{Free vector field localization}\label{ASubsec-5-1}

To start the discussion, we will consider the free field in a brane model with metric given by (\ref{2-01}). The Lagrangian for this case can be written as
\begin{eqnarray}\label{3-22}
\mathcal{L}^{m}(x,y)=-\frac{1}{4}\mathcal{F}_{MN}\mathcal{F}^{MN},
\end{eqnarray} 
with $\mathcal{F}_{MN}=\partial_{M}\mathcal{A}_{N}-\partial_{N}\mathcal{A}_{M}$. By calculating the equations of motion, we get
\begin{eqnarray}\label{3-23}
\partial_{M}\left[\sqrt{-g}\mathcal{F}^{MN}\right]=\partial_{\mu}\left[\sqrt{-g}\mathcal{F}^{\mu N}\right]+\partial_{k}\left[\sqrt{-g}\mathcal{F}^{kN}\right]=0.
\end{eqnarray}
Here, we can propose $\mathcal{A}_{N}=\left(\mathcal{A}^{T}_{\mu}+\partial_{\mu}\theta,\mathcal{B}_{k}\right)$ with $\partial^{\mu}\mathcal{A}^{T}_{\mu}=0$, and thus the equation (\ref{3-23}) can be split, for the components $\mathcal{A}^{T}_{\mu}$, as
\begin{eqnarray}\label{3-24}
\partial_{\mu}\left[\sqrt{-g}g^{\mu\nu}g^{\lambda\rho}\mathcal{F}^{T}_{\nu\lambda}\right]+\partial_{k}\left[\sqrt{-g}g^{kj}g^{\nu\rho}\partial_{j}\mathcal{A}^{T}_{\nu}\right]=0.
\end{eqnarray}
The equations of motion for the fields $\theta$ and $\mathcal{B}_{k}$ will not be important to our discussion, thus they will not be written here. From Eq. (\ref{3-24}) and by proposing $\mathcal{A}^{T}_{\mu}(x,y)=\hat{A}^{T}_{\mu}(x)\xi(y)$, we can get
\begin{eqnarray}
\frac{1}{\sqrt{-\hat{g}}}\partial_{\mu}\left[\sqrt{-\hat{g}}\hat{F}_{T}^{\mu\rho}(x)\right]=m^{2}\hat{A}_{T}^{\rho}(x),\label{3-25}\\
-\frac{e^{-(d-4)\sigma(y)}}{\sqrt{\Bar{g}}}\partial_{k}\left[\sqrt{\Bar{g}}\Bar{g}^{kj}e^{(d-2)\sigma(y)}\partial_{j}\xi(y)\right]=m^{2}\xi(y).\label{3-26}
\end{eqnarray}
Now, as well as we did for the scalar and spinor fields, the equation (\ref{3-26}) can be solved for $m^{2}=0$ and, with these zero-mode solutions, the vector field localization can be studied. Equation (\ref{3-26}) has a constant, $\xi_{0}(y)=c_{0}$, and also a non-constant solution for the zero-mode. This last one is closely related to the braneworld and its specific form is model dependent. Fortunately, in most cases, the constant zero-mode solution is the only one that can be confined \cite{Oda03,Oda04,Midodashvili,Freitas}. 

We are interested in studying the consistency of the confinement, therefore let us obtain the energy-momentum tensor for the Lagrangian (\ref{3-22}). By doing this, we get
\begin{eqnarray}\label{3-27}
\mathcal{T}^{m}_{MN}(x,y)=\mathcal{F}_{MP}(x,y)\mathcal{F}_{N}^{\ \ P}(x,y)+g_{MN}\mathcal{L}^{m}(x,y).
\end{eqnarray}
And, for the zero-mode, the components $\mathcal{T}^{m}_{\mu\nu}$ can be written as
\begin{eqnarray}\label{3-28}
\mathcal{T}^{m}_{\mu\nu}=e^{-2\sigma(y)}\xi^{2}_{0}(y)\left[\hat{F}^{T}_{\mu\rho}(x)\hat{F}_{\nu}^{T\ \rho}(x)+\hat{g}_{\mu\nu}(x)\hat{L}_{0}^{m}(x)\right].
\end{eqnarray}
From this, the consistency condition (\ref{2-14}) is satisfied only if $e^{-2\sigma(y)}\xi^{2}_{0}(y)=const$. As a first result, we get that the constant zero-mode solution cannot satisfy this requirement, and this is codimension independent. Therefore, for all those models where the confinement is performed with such solution, the localization is not consistent. For $5$D models, the free gauge field cannot be confined because the $K$ integral in (\ref{2-03}) is not finite \cite{Barut,Kehagias}. In this way, the result obtained from (\ref{3-28}) just confirms the non-localization of this field. However, most interesting results are obtained for codimension 2 and higher-dimensional models. In the literature, there are a large variety of models in these dimensional configurations where the metric (\ref{2-01}) gets the particular shape \cite{Gregory,Gherghetta,Carlos,Silva,Oda04}
\begin{eqnarray}\label{3-29}
ds^{2}=e^{2\sigma(r)}\hat{g}_{\mu\nu}(x)dx^{\mu}dx^{\nu}+dr^{2}+e^{2\sigma_{2}(r)}d\Omega^{2}_{n-1}.
\end{eqnarray}
Localization for free vector field in such scenarios was already studied \cite{Oda03,Oda05,Oda06,Midodashvili,Costa01,Costa02}. Generally, for all these references, the constant zero-mode solution is confined given that the $K$ integral, in (\ref{2-03}), is finite. However, by using our consistency conditions, we get that the localization of the free gauge field in such scenarios is not consistent. Therefore, even with a localized zero-mode, it is not possible to ignore back-reaction effects from the $U(1)$ gauge field on the backgorund geometry. Of course, there are still other braneworld models where the metric is not like that in (\ref{3-29}), as those in Refs. \cite{Choudhury01,Arkani02,Flachi}. However, the conclusion for these cases is the same: the constant zero-mode solution is not consistent with Einstein's equations. That way, the vector field localization really seems to need a mechanism to be consistent, and in the next section we  will discuss some of them.

\subsection{Vector field localization through mechanisms}\label{ASubsec-5-2}

Now, we will review some mechanisms used to confine the abelian vector field. Again, the focus is to verify the consistency of localization procedure with Einstein's equations. Let us first consider the codimension 1 braneworld models. In this dimensional configuration, the metric (\ref{2-01}) can be written as 
\begin{eqnarray}\label{3-30}
ds^{2}=e^{2\sigma(y)}\hat{g}_{\mu\nu}dx^{\mu}dx^{\nu}+e^{2\sigma_{2}(y)}dy^{2}.
\end{eqnarray}
With this, we can discuss the consistency for a wide variety of models, whether with thin or thick brane, with or without internal structure.\\

\noindent{\bf i) Scalar field coupling}\\

\noindent
Let us start by considering a localization mechanism commonly used to confine the spin $1$ field on thick branes \cite{Kehagias,Fu01,Chumbes,Cruz01}. In these models, the gauge field is coupled to some scalar function $G\left(y\right)$ through an action like the one given below
\begin{equation}\label{3-31}
S=-\frac{1}{4}\int d^{d}xdy\sqrt{-g}G\left(y\right)\mathcal{F}_{MN}\mathcal{F}^{MN}.
\end{equation}
The function $G\left(y\right)$ will be defined later for some known cases. By calculating the equation of motion and assuming the gauge $\mathcal{A}_{M}=(\mathcal{A}_{\mu},\mathcal{A}_{d+1}=0)$, we get the separated equations 
\begin{eqnarray}
 \frac{1}{\sqrt{-\hat{g}}}\partial_{\mu}\left[\sqrt{-\hat{g}}\hat{g}^{\mu\rho}\hat{g}^{\nu\lambda}\mathcal{F}_{\rho\lambda}\right]=m^{2}\hat{g}^{\nu\lambda}\hat{A}_{\lambda}(x),\label{3-32}\\
-\frac{e^{-(d-4)\sigma-\sigma_{2}}}{G}\partial_{y}\left[e^{(d-2)\sigma-\sigma_{2}}G\partial_{y}\xi\right]=m^{2}\xi,\label{3-33}
\end{eqnarray}
where, we already used the metric (\ref{3-30}) and also $\mathcal{A}_{\mu}(x,y)=\hat{A}_{\mu}(x)\xi(y)$. Beyond this, the effective action on the brane can be written as
\begin{eqnarray}\label{3-34}
S=-\frac{1}{4}\!\int\! d^{d}x\sqrt{-\hat{g}}\left[\hat{F}_{\mu\nu}\hat{F}^{\mu\nu}\!\!+\!2m^{2}\hat{A}_{\mu}\hat{A}^{\mu}\!\right]K.
\end{eqnarray}
where, Neumann boundary conditions was used for $\xi(y)$, and the quantity $K$ is given by
\begin{eqnarray}\label{3-35}
K=\int dye^{(d-4)\sigma(y)+\sigma_{2}(y)}G\left(y\right)\xi^{2}(y).
\end{eqnarray}
With this, by observing (\ref{3-34}), a gauge field (massless mode) on the brane can be obtained by doing $m^{2}=0$. Moreover, by properly choosing the function $G(y)$, this massless mode can be confined. In Refs. \cite{Kehagias,Fu01} this function is chosen in the form $G(y)=e^{\lambda\pi(y)}$, where $\pi(y)$ is a scalar field, namely the dilaton. From this, the zero-mode localization is obtained for some values $\lambda$. In Ref. \cite{Chumbes} the authors get a general form for $G(y)$, that should be valid for an arbitrary thick brane with $\sigma_{2}(y)=0$. In another direction, Ref. \cite{Zhao02} proposes $G(y)$ as a function of the Ricci scalar, and also for this case the localization can be obtained.

Let us turn back to the consistency conditions (\ref{2-14}) and (\ref{2-15}). From action (\ref{3-35}), the energy-momentum tensor can be written as
\begin{eqnarray}\label{3-36}
T_{MN}^{(\mbox{mat})}=G(y)\left[\mathcal{F}_{MP}\mathcal{F}_{N}^{\ \ P}-\frac{1}{4}g_{MN}\mathcal{F}_{PQ}\mathcal{F}^{PQ}\right].
\end{eqnarray}
Thus, by using the above configuration and considering only the zero-mode solution, we get
\begin{eqnarray}\label{3-37}
T_{\mu\nu}^{(\mbox{mat})}=G(y)e^{-2\sigma(y)}\xi_{0}^{2}(y)\left[\hat{F}_{\mu \rho}(x)\hat{F}_{\nu}^{\ \ \rho}(x)\right.\nonumber\\-\left.\frac{1}{4}\hat{g}_{\mu\nu}(x)\hat{F}_{\mu\nu}(x)\hat{F}^{\mu\nu}(x)\right].
\end{eqnarray}
where $\xi_{0}^{2}(y)$ is the zero-mode ($m^{2}=0$) solution of (\ref{3-33}). From expression (\ref{3-37}), we get that the consistency condition (\ref{2-14}) is satisfied only if $G(y)e^{-2\sigma(y)}\xi_{0}^{2}(y)=const.$. That way, by using the constant solution for the massless mode $\xi_{0}(y)$, we conclude that $G(y)$ only can assume the specific form $G(y)=e^{2\sigma(y)}$. The same conclusion is obtained from the condition (\ref{2-15}).

That condition on the function $G(y)$ considerably restricts the allowed models for this type of coupling. For example, in Ref. \cite{Kehagias}, the authors define $G(y)=e^{\frac{\lambda}{2}\sigma(y)}$. In this case, we get that the coupling constant $\lambda$ must be defined as $\lambda=4$, for the localization to be well-defined. In Ref. \cite{Fu01} is used $G(y)=e^{\tau \sqrt{3r}\sigma(y)}$, and the localization is obtained for $\tau\geq -\sqrt{\frac{r}{3}}$, with $0<r<1$; or $\tau>-\sqrt{\frac{1}{3r}}$, with $r>1$. By using our consistence condition, we get that $\tau=\frac{2}{\sqrt{3r}}$. Therefore, for both references localization can be done consistently. There is a very interesting reference, namely \cite{Chumbes}, where function $G(y)$ is defined as $G(y)=G(\phi)\propto \frac{\partial W^{2s}\left(\phi\right)}{\partial \phi}$, with $\underline{s}$ a constant, and $W(\phi)$ the superpotential related to the scalar field $\phi(y)$ which generate the braneworld. The authors show that for a brane model generated by scalar field with Sine-Gordon potential \cite{Gremm01}, i.e.,
\begin{eqnarray}\label{3-38}
W(\phi)\!=\!3bc\sin\!\left[\!\sqrt{\frac{2}{3b}}\phi\right],\ \mbox{with}\ \phi(y)=\sqrt{6b}\arctan\left(cy\right),
\end{eqnarray}
function $G(\phi)$ is given by $G(y)=\mbox{sech}^{2s}\left(2cy\right)$ and localization can be obtained. When we compare this with our result, which for the specific model \cite{Gremm01} stays $G(y)=e^{2\sigma(y)}=\mbox{sech}^{2b}\left(2cy\right)$, the constant $s$ should be $s=b$. On the other hand, for the brane model \cite{Kehagias}, where
\begin{eqnarray}\label{3-39}
W(\phi)=av\left(\phi-\frac{\phi^{3}}{3v^{2}}\right),\ \ \mbox{with}\ \ \phi(y)=v\arctan\left(ay\right),
\end{eqnarray}
the Ref. \cite{Chumbes} obtain $G(y)=\mbox{sech}^{4s}\left(ay\right)$. Now, by using our result $G(y)=e^{2\sigma(y)}=\mbox{sech}^{4b}\left(ay\right)e^{-b\tanh^{2}(ay)}$, we see that functions $G(y)$ does not match, and the superpotential (\ref{3-39}) does not allow a consistent localization. Similar result can be obtained for deformed thick brane models \cite{Bazeia05}. This conclusion indicates that $G(y)=G(\phi)\propto \frac{\partial W^{2s}\left(\phi\right)}{\partial \phi}$ have not a general validity as localization mechanism, i.e., it does not work for any braneworld model. Therefore, except for (\ref{3-38}), the function $G(y)=G(\phi)\propto \frac{\partial W^{2s}\left(\phi\right)}{\partial \phi}$ does not provide a consistent localization for the gauge field (zero-mode). Another interesting model is presented in Ref. \cite{Zhao02}, where $G(y)$ is function of the Ricci scalar, namely,
\begin{equation}\label{3-31a}
S=-\frac{1}{4}\int d^{4}xdy\sqrt{-g}G\left(R\right)\mathcal{F}_{MN}\mathcal{F}^{MN}.
\end{equation}
The authors argue that, if $G(R)$ is a continuous function, the zero-mode localization of the vector field is determined by the behavior of $G(R)$ when $y\to\infty$. They get that this function must be, asymptotically, something like $G(R_{\infty})\propto |y|^{-p}$, with $p$ a positive value. Considering the AdS feature of the space, the authors show that the warp factor must be, asymptotically, in the shape
\begin{eqnarray}
e^{2\sigma(y\to\pm\infty)}\to |y|^{-2}.
\end{eqnarray}
Therefore, the consistency condition obtained by us for models like (\ref{3-31}), i.e., $G(y)=e^{2\sigma(y)}$, can be satisfied for $G(R_{\infty})\propto |y|^{-p}$, if $p=2$. However, we cannot say that this is valid for another range of the variable $y$. Moreover, there seems to be a contradiction in the arguments used by the authors themselves. They propose a localization mechanism in an asymptotically AdS space-time, thus, the Ricci scalar is, in that limit, $R(|y|\to\infty)\propto -C_{R}$ (constant). Therefore, $G(R)$ should go to a constant value at that limit, and the localization cannot be reached. Anyway, the requirement of $G(R)=e^{2\sigma(y)}$ does not seem so easy to meet for an arbitrary model. Other interesting points can also be discussed. For example, for models like \cite{Kehagias,Fu01}, we obtain that the coupling parameter ($\lambda$ or $\tau$) are not free, they must be fixed by consistency reasons. In this way, the analysis performed in Refs. \cite{Landim01,Landim02,Landim03} by research resonances of the gauge field with action like (\ref{3-35}) should be reevaluated. Since, there is no freedom in choosing the parameters $\lambda$ or $\tau$, used to plot the graphics in those references. \\

\noindent{\bf ii) G-N localization mechanism}\\

\noindent
Now, let us verify the non-covariant mechanism proposed by K. Ghoroku and A. Nakamura (G-N) in Ref. \cite{Ghoroku}. In this paper is used a metric like (\ref{3-30}) with the warp factors given by $\sigma_{2}(y)=\sigma(y)=-\ln\left(1+k|y|\right)$. The Lagrangian for the vector field with G-N mechanism is written as
\begin{equation}\label{3-40}
\mathcal{L}^{m}=-\frac{1}{4}\mathcal{F}_{MN}\mathcal{F}^{MN}-\frac{1}{2}\left[M^{2}+c\delta(y)\right]\mathcal{A}_{M}\mathcal{A}^{M}.
\end{equation}
This model, although not being gauge invariant or even covariant, the effective theory on the brane has the desired features: a massless vector field with gauge symmetry. After some steps like those performed in the previous case, we can get an EOM for the effective vector field $\mathcal{A}^{T}_{\mu}$, and by proposing the separation of variables $\mathcal{A}^{T}_{\mu}=\hat{A}^{T}_{\mu}(x)\xi(y)$, the localization discussion can be carried out. By doing this, the zero-mode solution can be obtained with the ansatz 
\begin{eqnarray}\label{3-41}
\xi_{0}(y)=c_{0}e^{a\sigma(y)},\ \mbox{where}\ a=-\frac{c}{2k}=\frac{1}{k}\left(\!\sqrt{k^{2}\!+\!M^{2}}\!-\!k\!\right),
\end{eqnarray}
and it will be localized if $a>0$. With this, we can analyze the consistency conditions (\ref{2-14}) and (\ref{2-15}) for the energy-momentum tensor. From the Lagrangian (\ref{3-40}), we get for the zero-mode
\begin{eqnarray}\label{3-42}
T_{\mu\nu}^{(\mbox{mat})}(x,y)=e^{-2\sigma}\xi^{2}_{0}\left[\hat{F}^{T}_{\mu\rho}(x)\hat{F}_{\nu}^{T\ \rho}(x)\right.\nonumber\\\left.-\frac{1}{4}\hat{g}_{\mu\nu}(x)\hat{F}^{T}_{\rho\alpha}(x)\hat{F}_{T}^{\rho\alpha}(x)\right].
\end{eqnarray}
Thus, by using the zero-mode solution, we get that $e^{-2\sigma}\xi^{2}_{0}=e^{2(a-1)\sigma}=const.$ and the consistency condition (\ref{2-14}) will be satisfied when $a=1$. This value of $a$ fix all parameters in the Lagrangian (\ref{3-40}) in the following shape, $c=-2k$ and $|M|=\sqrt{3}|k|$. By a similar analysis, we show that consistency condition (\ref{2-15}) gives us the same result and this localization mechanism can be performed consistently.\\

\noindent{\bf iii) Non-minimal coupling with gravity}\\

\noindent
Finally, let us discuss the localization mechanism proposed in Refs. \cite{Flachi,Zhao,Alencar02}. In these works, the G-N mechanism is used as a motivation to propose the vector field coupling with gravity through the scalar and the Ricci tensor. The action for this case is given by
\begin{eqnarray}\label{3-43}
S=\frac{1}{2}\int d^{d}xdy\sqrt{-g}\left[-\frac{1}{2}\mathcal{F}_{MN}\mathcal{F}^{MN}+\lambda_{1}R\mathcal{A}_{M}\mathcal{A}^{M}\right.\nonumber\\+\left.\lambda_{2}R_{MN}\mathcal{A}^{M}\mathcal{A}^{N}\right].
\end{eqnarray}
Where $\underline{d}$ is the brane dimension, $R$ and $R_{MN}$ are the scalar and the Ricci tensor, respectively. By proposing again $\mathcal{A}_{M}=\left(\mathcal{A}^{T}_{\mu}+\partial_{\mu}\theta,\mathcal{B}\right)$, and after some steps, it is possible to obtain an EOM for the transverse field $\mathcal{A}^{T}_{\mu}$. In order that, by using the separation of variables $\mathcal{A}^{T}_{\mu}(x,y)=\hat{A}^{T}_{\mu}(x)\xi(y)$, we obtain a zero-mode solution for $\xi(y)$ given by
\begin{eqnarray}\label{3-44}
\xi_{0}(y)=e^{a\sigma(y)},\ \mbox{with}\ a=-2\lambda_{1}\left(D-1\right)-\lambda_{2}.
\end{eqnarray}
Beyond the additional conditions
\begin{eqnarray}
\left[\frac{D-4}{2}+a\right]^{2}&=&\frac{\left(D-4\right)^{2}}{4}-\left(\lambda_{1} \left(D-1\right)+\lambda_{2}\right)(D-2),\label{3-45}\\
\frac{D-4}{2}+a &>&\frac{1}{2},\label{3-46}
\end{eqnarray}
where $D=d+1$. The conditions, (\ref{3-44}) and (\ref{3-45}), are required for a zero-mode solution to exist, and the condition (\ref{3-46}) should be satisfied for that the solution $\xi_{0}$ be confined on the brane. By doing a similar analysis like that in (\ref{3-42}), we conclude that the consistency conditions (\ref{2-14}) and (\ref{2-15}) can be satisfied for $a=1$. Therefore, this localization mechanism can provide a confinement of the vector field consistent.

By eliminating higher-order terms, the brane components of energy-momentum tensor for the action (\ref{3-43}) can be written, for the zero-mode, as
\begin{eqnarray}\label{3-47}
T_{\mu\nu}^{(\mbox{mat})}(x,y)=e^{-2\sigma}\xi^{2}_{0}\left[\hat{F}^{T}_{\mu\rho}(x)\hat{F}_{\nu}^{T\ \rho}(x)+\hat{g}_{\mu\nu}(x)L_{0}(x)\right]\nonumber\\
=e^{2(a-1)\sigma}\left[\hat{F}^{T}_{\mu\rho}(x)\hat{F}_{\nu}^{T\ \rho}(x)+\hat{g}_{\mu\nu}(x)L_{0}(x)\right].
\end{eqnarray}
Thus, we see that the condition (\ref{2-14}) is satisfied when $a=1$. When we use $a=1$, the localization condition (\ref{3-46}) can be written as $d-2>0$, and it is always satisfied for models with $d\geq 4$. By doing $a=1$ in Eqs. (\ref{3-44}) and (\ref{3-45}), we get 
\begin{eqnarray}
\lambda_{1}=-\frac{1}{(D-2)(D-1)}.\label{3-48}
\end{eqnarray}
This result shows that vector field localization, by using the mechanism (\ref{3-43}), is consistent with Einstein equation only when both interaction terms are switched on simultaneously. Beyond this, the two parameters $\lambda_{1}$ and $\lambda_{2}$ are completely fixed by consistency reasons. This result allows us to comment briefly that one presented in Ref. \cite{Sui}, where the authors plot some graphics of the potential and the relative probability for various values of $a$. As we obtained from the consistency conditions, the parameters are fixed and such freedom for the parameter $a$ does not exist. In fact, the authors argue that massive resonant modes can exist if $a>3$, which, when compared with our results $a=1$, shows that resonant modes cannot exist.\\

\noindent{\bf iv) Localization in codimension $2$ or higher models}\\

\noindent
Generally, for most of the models in co-dimendion $2$ or higher, the free $U(1)$ gauge field is already naturally confined just by minimum couplings with gravity \cite{Oda03,Oda04,Oda06,Midodashvili,Costa02,Torrealba,Giovannini,Fu02}. Thus, there are not many localization mechanisms for this field in those dimensional configurations. However, as we saw in subsection (\ref{ASubsec-5-1}), the free field case is already not consistent with Einstein's equations, so some localization mechanism really seems to be necessary. In Ref. \cite{Flachi}, the vector field is confined in codimension 2 intersecting delta-like branes by proposing a mechanism like that in equation (\ref{3-43}). In Ref. \cite{Freitas}, this study is performed for a generic model with arbitrary codimension embedded in asymptotically AdS space. For both cases, the results are similar to these obtained early in item (iii). In other words, the consistency with Einstein's equations is obtained just when both interaction terms are switched on simultaneously. In fact, this conclusion is codimension independent for this localization mechanism. Let the action for the vector field in a generic model with the localization mechanism like (\ref{3-43}) be given by
\begin{eqnarray}\label{3-51}
S=\frac{1}{2}\int d^{d}xdy^{n}\sqrt{-g}\left[-\frac{1}{2}\mathcal{F}_{MN}\mathcal{F}^{MN}+\lambda_{1}R\mathcal{A}_{M}\mathcal{A}^{M}\right.\nonumber\\+\left.\lambda_{2}R_{MN}\mathcal{A}^{M}\mathcal{A}^{N}\right],
\end{eqnarray}
where, $\underline{d}$ is the brane dimension and $\underline{n}$ is the codimension. By performing some steps like that in the item (iii), we can get a zero-mode solution given by
\begin{eqnarray}\label{3-52}
\xi_{0}(y)=e^{a\sigma(y)},\ \mbox{with}\ a=-2\lambda_{1}\left(D-1\right)-\lambda_{2}.
\end{eqnarray}
With the additional conditions
\begin{eqnarray}
\left[\frac{D-4}{2}+a\right]^{2}&=&\frac{\left(D-4\right)^{2}}{4}-\left(\lambda_{1} \left(D-1\right)+\lambda_{2}\right)(D-2),\label{3-53}\\
\frac{D-4}{2}+a &>&\frac{n}{2},\label{3-54}
\end{eqnarray}
where $D=d+n$. Again, to verify the consistency with Einstein's equations, we need the energy-momentum tensor. By eliminating higher order terms, the brane components of energy-momentum tensor, for the action (\ref{3-51}), can be written for the zero-mode as,
\begin{eqnarray}\label{3-55}
T_{\mu\nu}^{(\mbox{mat})}(x,y)=e^{-2\sigma}\xi^{2}_{0}\left[\hat{F}^{T}_{\mu\rho}(x)\hat{F}_{\nu}^{T\ \rho}(x)+\hat{g}_{\mu\nu}(x)L_{0}(x)\right]\nonumber\\
=e^{2(a-1)\sigma}\left[\hat{F}^{T}_{\mu\rho}(x)\hat{F}_{\nu}^{T\ \rho}(x)+\hat{g}_{\mu\nu}(x)L_{0}(x)\right].
\end{eqnarray}
Thus, we see that the condition (\ref{2-14}) is satisfied when $a=1$. If we use $a=1$, the localization condition (\ref{3-54}) can be written as $d-2>0$, which is always satisfied for models with $d\geq 4$. By doing $a=1$ in Eqs. (\ref{3-52}) and (\ref{3-53}), we get
\begin{eqnarray}
\lambda_{1}=-\frac{1}{(D-2)(D-1)}.\label{3-56}
\end{eqnarray}
Therefore, again, we get the same conclusion that presented in the item (iii): the localization mechanism (\ref{3-51}) is consistent with Einstein's equations only when both interaction terms are switched on simultaneously. In this way, any back-reaction effect from the vector field on the background metric can be eliminated, at least for the zero-mode.

\section{Final Remarks}\label{ASec-6}

In this work, we discussed the consistency of fields localization in braneworld models. By studying Einstein's equations in the presence of matter fields, we obtained the constraints (\ref{2-14}) and (\ref{2-15}) for the energy-momentum tensor that should be valid for any brane model. Such constraints are a consequence of the assumption used in fields localization, where a confined matter field does not modify the bulk metric. In this way, the localization procedure used in the literature will be consistent only if such conditions are satisfied. We applied these consistency conditions for some cases, namely, the spin $0$, $\frac{1}{2}$ and $1$ Standard Model fields, with and without localization mechanisms.

For the scalar field, as discussed in many references \cite{Barut,Kehagias,Oda02,Oda03,Liu01,Liu02,Fu01,Midodashvili}, there is a constant zero-mode solution that can be localized. By using this confined constant solution, we showed in Eqs. (\ref{3-06}) and (\ref{3-07}) that the energy-momentum tensor for the scalar field satisfies the consistency conditions (\ref{2-14}) and (\ref{2-15}). Therefore, the scalar field localization (zero-mode) is consistent with Einstein's equations. After, we apply those conditions for the spin $\frac{1}{2}$ field in codimension 1 models, with a Yukawa-like interaction given by $\mathcal{L}^{m}_{in}=\lambda f(y)\Bar{\Psi}\Psi$. Also, for this field, there is a variety of models with this kind of coupling \cite{Koley01,Kehagias,Melfo,Castro,Mendes,Ringeval}. For each of these cases, a different Yukawa interaction is proposed and the spinor zero-mode localization, actually one of the chiralities, is obtained for some properly condition. About consistency conditions (\ref{2-14}) and (\ref{2-15}), we obtained the energy-momentum tensor (\ref{3-20}). And from this, the spinor field localization is consistent with Einstein's equations only if $f(y)\propto e^{-\sigma_{2}(y)}\sigma_{1}'(y)$, with $\sigma_{1}$ and $\sigma_{2}$ the warp factors in (\ref{3-08}). As discussed in section (\ref{ASec-4}), this relation eliminates the freedom to choose the function $f(y)$. With this, the Yukawa interaction used in Refs. \cite{Barut, Oda04}, for RS-II type braneworlds, is consistent with the Einstein's equations by properly choose the interacting parameter $\lambda$. However, for those functions $f(y)$ used in thick brane models like \cite{Koley01,Kehagias,Melfo,Castro,Mendes,Ringeval}, the localization is not consistent and it should be reviewed. We must to stress that the analysis performed for the spinor field considered only Yukawa-like interactions, and also only in codimension 1 models. There are still other localization mechanisms and other dimensional configurations where this analysis could be carried out as for exemple \cite{GAlencar}.

Finally, we verified the consistency condition for the vector field. As discussed widely in the literature \cite{Barut,Oda03,Oda04,Midodashvili,Freitas}, the free gauge field (zero-mode) cannot be confined in $5$D, however, for some higher codimension models it can be localized. In section (\ref{ASubsec-5-1}), we obtained the energy-momentum tensor (\ref{3-27}) for the gauge field and, from this, the consistency conditions were analyzed. As a general result, the conditions (\ref{2-14}) and (\ref{2-15}) are consistent for the zero-mode vector field only if $e^{-2\sigma(y)}\xi^{2}_{0}(y)$ is a constant. However, as discussed in that section, there is not zero-mode solution $\xi_{0}$ confined that satisfies the above condition. Such result is independent of the braneworld model or the number of extra dimension, thus the localization of this field is not consistent with Einstein's equations and a mechanism to confine it really seems necessary. In this direction, we analyzed the consistency of some localization mechanisms in section (\ref{ASubsec-5-2}). For example, that mechanism proposed in Refs. \cite{Kehagias,Fu01,Chumbes,Zhao02}, where the action for the gauge field is given by something like (\ref{3-31}). For these kinds of coupling, there is a zero-mode constant solution that can always be localized when $G(y)$ is like Gaussian. By using the consistency conditions, the localization of $\xi_{0}=c_{0}$ is consistent with Einstein's equations just if $G(y)=e^{2\sigma(y)}$ in Eq. (\ref{3-37}). In this way, the Gaussian feature is confirmed, however, such expression does not present any free coupling parameter. Beyond this, that function $G(y)$ eliminates some mechanism proposed in the literature, like that in Ref. \cite{Zhao02}, where $G(y)=G(R)$ is a function of Ricci scalar. Other interesting points can still be discussed here. As there is not a free parameter in $G(y)=e^{2\sigma(y)}$, the analysis performed in Refs. \cite{Landim01,Landim02,Landim03} about resonant modes of the gauge field with action given by (\ref{3-31}), for models like \cite{Kehagias,Fu01}, should be reevaluated. Since some results are obtained by using a free coupling parameter, which, by our analysis, does not exist. We also discussed the non-covariant mechanism proposed in Ref. \cite{Ghoroku}. For this case, the zero-mode solution for the gauge field sector is given by $\xi_{0}(y)=e^{a\sigma(y)}$, and the consistency conditions (\ref{2-14}) and (\ref{2-15}) were satisfied when $a=1$. With this value of $a$, all parameters in (\ref{3-40}) were fixed, namely, $c=-2k$ and $|M|=\sqrt{3}|k|$. In Ref. \cite{Ghoroku} is also discussed the localization of the scalar component, and the authors conclude that both sectors cannot be confined simultaneously. Moreover, the theory does not indicate what these sectors should be confined. Maybe, the consistency condition could be used to solve this, but we did not perform such study here. Inspired by this mechanism, we analyzed the localization mechanism proposed in Refs. \cite{Zhao,Alencar02}. And by starting from (\ref{3-43}), the consistency with Einstein's equations was obtained only if both interaction terms are present. And just like in the previous cases, the parameters $\lambda_{1}$ and $\lambda_{2}$ in (\ref{3-43}) were completely fixed by consistency reasons. Beyond all these codimension 1, higher codimension models were investigated. And also for this dimensional configuration, by using the localization mechanism (\ref{3-51}), the consistency with Einstein's equations is possible only if both interaction terms are switched on simultaneously. In this way, we believe that a comprehensive analysis was performed about the consistency of fields localization, and such study can be used as a guide for building new confining mechanisms.

\vfill
\section*{Acknowledgment}
The authors would like to thank Alexandra Elbakyan and sci-hub, for removing all barriers in the way of science. We acknowledge the financial support provided by the Conselho Nacional de Desenvolvimento Científico e Tecnológico (CNPq) and Fundação Cearense de Apoio ao Desenvolvimento Científico e Tecnológico (FUNCAP) through PRONEM PNE-0112-00085.01.00/16.

\end{document}